\documentstyle[aps,pra,epsf,multicol]{revtex}
\begin{document}
\draft
\title{Fundamental bounds on quantum measurements with a mixed apparatus}
\author{S. Bose  and  V. Vedral}
\address{Centre for Quantum Computation, Clarendon Laboratory,
	University of Oxford,
	Parks Road,
	Oxford OX1 3PU, England}

\maketitle
\begin{abstract}
We consider the apparatus in a quantum measurement process to be
in a mixed state. We propose a simple upper bound on the probability
of correctly distinguishing any number of mixed states.
We use this to derive fundamental bounds on the efficiency of a
measurement in terms of the temperature of the apparatus.  
\end{abstract}

\pacs{Pacs No: 03.65.Bz, 03.67.-a}

\begin{multicols}{2}
The quantum measurement process was first analysed from the
fully quantum perspective by von Neumann \cite{everett}. 
In the description of a typical quantum measurement, the initial states of the system and the apparatus (both treated quantum mechanically) are taken 
to be pure states. The measurement process is a transformation of
the type

\begin{eqnarray}
\int \phi(x) |x\rangle_{\scriptsize s} dx &\otimes& \int \eta(y) |y\rangle_{\scriptsize a} dy \nonumber \\
&\rightarrow&  \int \int \phi(x) \eta(y+f(x))
|x\rangle_{\scriptsize s} \otimes |y\rangle_{\scriptsize a} dx dy
\end{eqnarray}
where $\int \phi(x) |x\rangle_{\scriptsize s} dx$ and $\int \eta(y) |y\rangle_{\scriptsize a} dy$ are initial states of the system and the
measuring apparatus respectively. Note that after the measurement, the
state of the apparatus is correlated to the state of the system. This
enables us to infer the state of the system by observing the state
of the apparatus. In general, one is allowed to perform {\em any} 
positive operator valued measurement (POVM) on the apparatus state
to infer the system state (not analysed by von Neumann). This becomes
important in a realistic quantum measurement, in which the
apparatus is a macroscopic system and likely to be in a
mixed (generally thermal) state throughout the measurement. This will
make measurements more difficult in general than the pure apparatus
case as mixed states are generically harder to distinguish. One could,
of course, attempt to estimate the efficiency of state inference on the
basis of specific von Neumann projections on the apparatus state. This, however,
is usually less efficient than a more general POVM. Strangely,
to the best of our knowledge, all treatments of quantum measurement
seem to neglect this fact and assume the initial apparatus state to
be pure. In this letter, we analyse the quantum measurement process
with a mixed apparatus. Our formalism allows us to estimate the 
probability of success in state determination with a mixed apparatus.
We use this to put bounds on the probability os successful
 state inference
in terms of the temperature of an apparatus in a thermal state.  

   We start with a simple example of a two level system being measured
by a harmonic oscillator apparatus initially in a thermal state. The
measurement interaction is described by
\begin{eqnarray}
|0\rangle_{\scriptsize s}\otimes |n\rangle_{\scriptsize a} &\rightarrow&  
|0\rangle_{\scriptsize s}\otimes |n\rangle_{\scriptsize a} \nonumber \\
|1\rangle_{\scriptsize s}\otimes |n\rangle_{\scriptsize a} &\rightarrow&  
|1\rangle_{\scriptsize s}\otimes |n+1\rangle_{\scriptsize a},
\end{eqnarray}
where $|0\rangle_{\scriptsize s},|1\rangle_{\scriptsize s}$ are two orthogonal
states of the two level system that the apparatus is designed to detect and
$|n\rangle_{\scriptsize a}$ denotes a Fock state of the apparatus. Note that
we could have chosen any other measurement interaction, but that would
lead to similar results. In particular, this measurement interaction
works perfectly
when the apparatus is initially in a pure Fock state. If the apparatus
starts off in the initial thermal state $\rho(\beta)_{\scriptsize a}=\sum_{n=0}^{\infty} 
\frac{e^{-\beta n}}{Z}(|n\rangle \langle n|)_{\scriptsize a}$ (here $Z$ is the
partition function $\sum_{n=0}^{\infty} e^{-\beta n}$ and
$\beta=\hbar \omega/k_{\scriptsize B} T$ where $\omega$ is the frequency
of the oscillator, $T$ its temperature and $k_{\scriptsize B}$ is Boltzmann's constant), then the measurement leads to 
\begin{eqnarray}
(|0\rangle \langle 0|)_{\scriptsize s}\otimes \rho(\beta)_{\scriptsize a} &\rightarrow&  
(|0\rangle \langle 0|)_{\scriptsize s}\otimes \rho(\beta)_{\scriptsize a} \nonumber \\
(|1\rangle \langle 1|)_{\scriptsize s}\otimes \rho(\beta)_{\scriptsize a} &\rightarrow&  
(|1\rangle \langle 1|)_{\scriptsize s}\otimes \sum_{n=1}^{\infty} 
\frac{e^{-\beta (n-1)}}{Z}(|n\rangle \langle n|)_{\scriptsize a}.
\end{eqnarray}
The maximum probability $P_{\scriptsize c}$ of  correctly distinguishing between any two
mixed states $\rho_0$ and $\rho_1$ by any POVM is given by Helstrom's formula
\cite{hels}
\begin{equation}
P_{\scriptsize c}(\rho_0,\rho_1)= \frac{1}{2}+\frac{1}{4}\mbox{Tr}|\rho_0 - \rho_1|.
\end{equation}
In our case $\rho_0=\rho(\beta)_{\scriptsize a}$ and $\rho_1=\sum_{n=1}^{\infty} 
\frac{e^{-\beta (n-1)}}{Z}(|n\rangle \langle n|)_{\scriptsize a}$. We thus
obtain the probability of correctly identifying the state of the apparatus
(and hence the system, which is our main goal) to be
\begin{equation}
P_{\scriptsize c}= \frac{1}{2}+\frac{1}{4}\{\frac{1}{Z}+
\frac{e^{\beta}-1}{Z} (Z-1)\}.
\end{equation}

\begin{figure} 
\begin{center}
\leavevmode 
\epsfxsize=8cm 
\epsfbox{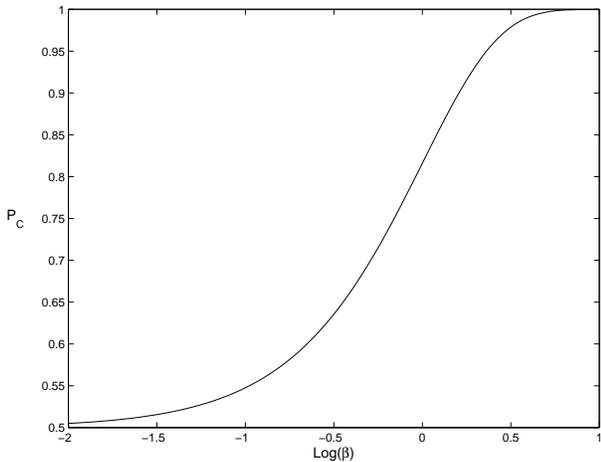}
\caption{\narrowtext The figure shows the variation of the probability of correct 
state inference of a two level system being measured by a harmonic oscillator
apparatus with $\log{\beta}$, where $\beta=\hbar\omega/k_{\scriptsize B} T$
and the logarithm is in base $10$.
Here $P_{\scriptsize c}$ is obtained from Helstrom's formula.}
\label{prob1}
\end{center}
\end{figure}
We see from fig.\ref{prob1} that we obtain $P_c\rightarrow 1$ for
$\beta\rightarrow \infty$ (low temperature limit) and $P_c \rightarrow 0.5$ for $\beta\rightarrow
0$ (high temperature limit). This means that at low temperatures we can correctly distinguish
between $|0\rangle_{\scriptsize s}$ and $|1\rangle_{\scriptsize s}$ and this
is because then the initial state of the apparatus is virtually pure. On the other hand for high temperatures, the initial state of the apparatus is
virtually maximally mixed and does not change due to the measurement
interaction (i.e $\rho_0=\rho_1$). We see that for achieving
$P_c$ greater than $0.8$, we require

\begin{equation}
\frac{\omega}{T} \geq \frac{k_{\scriptsize B}}{\hbar}.
\end{equation}
This can be seen as a fundamental limit on obtaining one bit of information
reliably.
  
   The quantum system being measured, may, however, be a system with more than
two orthogonal states. In that case we have to have a formula for the
probability of correctly identifying any one of several mixed states
$\rho_0,\rho_1,...,\rho_N$. There is no existing general formula extending the
Helstrom's formula to an arbitrary number of density matrices. Here we propose
a generalisation which gives a bound
on the probability of correct identification of one of $\rho_0,\rho_1,...,\rho_N$ which appear with probabilities 
$p_0,p_1,...,p_N$.  This bound is

\begin{equation}
\label{QIHG}
P_{\scriptsize c}(\rho_0,\rho_1,...,\rho_N;p_0,p_1,...,p_N)= e^{H-h(p_0,p_1,...,p_N)},
\end{equation}
where
\begin{equation}
H=S(\sum_i p_i \rho_i)-\sum_i p_i S(\rho_i),
\end{equation}
is the Holevo bound \cite{hol}, $S(\rho)=-\mbox{Tr} \rho \ln{\rho}$ and
where 
\begin{equation}
h(p_0,p_1,...,p_N) = -\sum_i p_i \ln{p_i},
\end{equation}
is the Shannon entropy of the probability distribution of the
density matrices. The rationale behind this formula becomes clear
when one considers a sequence of $n$ preparations of the system state and
measurements on the corresponding apparatus states. The probability
of correctly inferring a certain sequence of $n$ states $\{\rho_i\}$  is
bounded above by the ratio of the number of correctly identified
sequences and the total number of possible sequences. From the statistical
interpretation of the quantum relative entropy \cite{stat}, we get the
numerator of this ratio to be $e^{nH}$ and the denominator is
$e^{nh(p_0,p_1,...,p_N)}$ (this is equivalent to the law of
large numbers). We, in fact, use the formula for $n=1$ and hence the
Eq.(\ref{QIHG}) for $P_{\scriptsize c}$.

\begin{figure} 
\begin{center}
\leavevmode 
\epsfxsize=8cm 
\epsfbox{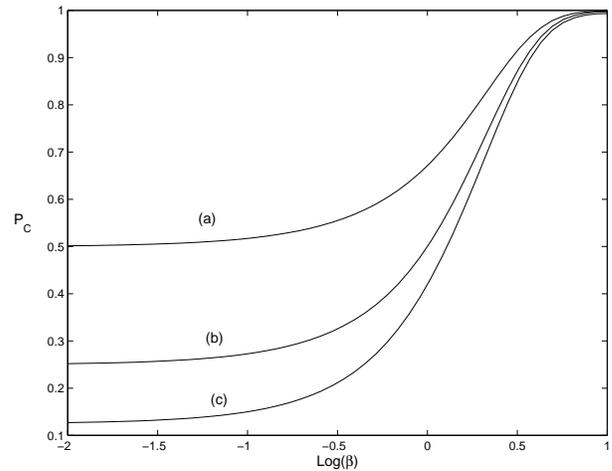}
\vspace*{1.0cm}
\caption{\narrowtext The figure shows the variation of the upper bound on the probability of correct 
state inference of a $N+1$ level system being measured by a harmonic oscillator
apparatus with $\log{\beta}$, where $\beta=\hbar\omega/k_{\scriptsize B} T$
and the logarithm is in base $10$.
For plot(a) $N=1$, for plot(b) $N=3$, for plot(c) $N=7$. Here $P_{\scriptsize c}$ is obtained from our formula Eq.(\ref{QIHG}).}
\label{prob2}
\end{center}
\end{figure}

We again take a harmonic oscillator apparatus,
but now a $N+1$ level system with orthogonal
states $|0\rangle_{\scriptsize s},|1\rangle_{\scriptsize s},...,|N\rangle_{\scriptsize s}$ which interacts with the apparatus
as
\begin{equation}
|i\rangle_{\scriptsize s}\otimes |n\rangle_{\scriptsize a} \rightarrow  
|i\rangle_{\scriptsize s}\otimes |n+i\rangle_{\scriptsize a}.
\end{equation}
The initial state of the apparatus is again a thermal state. Corresponding
to each system state $|i\rangle_{\scriptsize s}$, the apparatus will
evolve to a different state $\rho_{\scriptsize a}^i$. These are
given by
\begin{equation}
\rho_{\scriptsize a}^i=\sum_{n=i}^{\infty} 
\frac{e^{-\beta (n-i)}}{Z}(|n\rangle \langle n|)_{\scriptsize a}.
\end{equation}
The entropy of all these states is the same and equal to
\begin{equation}
S(\rho_{\scriptsize a}^i)=\frac{\langle E \rangle}{k_{\scriptsize B} T}
+\ln{Z},
\end{equation}
where $\langle E \rangle=(\hbar\omega/Z)\sum n e^{-\beta n}$ is the average energy of the apparatus.

The entropy of the total state $\rho=\sum_i \rho_{\scriptsize a}^i$ is given by
\begin{eqnarray}
S(\rho)&=&-\sum_{i=0}^N (\sum_{j=0}^i \frac{e^{-\beta j}}{Z(N+1)})\ln{(\sum_{j=0}^i\frac{e^{-\beta j}}{Z(N+1)})} \nonumber \\&-&
\sum_{i=N+1}^\infty (\sum_{j=i-N}^i \frac{e^{-\beta j}}{Z(N+1)})\ln{(\sum_{j=i-N}^i\frac{e^{-\beta j}}{Z(N+1)})}
\end{eqnarray}
The entropies $S(\rho)$ and $S(\rho_{\scriptsize a}^i)$ are used
to compute $P_{\scriptsize c}$ from Eq.(\ref{QIHG}). This is plotted in
Fig. \ref{prob2} for three different dimensionalities of the measured
system ($N+1=2,4,8$). Note that all the three plots in Fig.\ref{prob2}
have the same shape and, as expected, tend to $\frac{1}{N+1}$ for
small $\beta$ (high temperature limit) and to unity for large
$\beta$ (low temperature limit). Note also that in the latter case 
(when $\beta > 1$) the
probability bound $P_c$ is well approximated by the simple
expression

\begin{equation}
\label{appr}
P_c \sim \frac{e^{-\frac{\langle E \rangle}{k_{\scriptsize B} T}}}{Z}.
\end{equation}
This is because, at low temperatures, $S(\rho) \sim h(p_0,p_1,...,p_N)$,
so that $P_c \sim e^{-S(\rho_{\scriptsize a}^0)}$. The above approximation
(Eq.(\ref{appr})) is already very good for $\beta=5$ differing by about $2$ percent from the exact value.

       Our bound applies to the general setting of measuring states
of a system by correlating them to pure nonorthogonal apparatus states.
This happens when different orthogonal states of the system get correlated
to different nonorthogonal states of the apparatus (for example,
when different Fock states inside
a cavity are inferred by different dispacements of a mirror in a coherent state
\cite{cavm}). In quantum optics, in particular, 
preparation of nonclassical states via conditional measurements on an
apparatus is very popular \cite{kurt}. Our formula for $P_{\scriptsize c}$
will be an upper bound to the fidelity of such preparations.
We hope that this kind of analysis also stimulates more research in 
the area of quantum computation with mixed states as explored
in Ref.\cite{ccqc}.

     \end{multicols}
\end{document}